\documentclass[a4paper]{jpconf}
\usepackage{graphicx,wrapfig}

\newcommand{\alp}{\ensuremath{\alpha}}
\newcommand{\ardm}{ArDM}
\newcommand{\mgcs}{\ensuremath{\,\mathrm{mg/cm^{2}}}}
\newcommand{\kevr}{\ensuremath{\,\mathrm{keV_{r}}}}
\newcommand{\keve}{\ensuremath{\,\mathrm{keV_{ee}}}}

\begin{document}
\title{The Argon Dark Matter Experiment (\ardm )}

\author{Christian Regenfus}

\address{Physik-Institut der Universit¬\"at Z¬\"urich, CH--8057 Z¬\"urich, Switzerland\\
\vspace{1mm}on behalf of the \ardm\ collaboration}

\ead{regenfus@cern.ch}

\begin{abstract}

The \ardm\ experiment, a 1\,ton liquid argon TPC/Calorimeter, is designed for the 
detection of dark matter particles which can scatter off the spinless argon 
nucleus, producing nuclear recoils. These events will be discerned by their 
light to charge ratio, as well as the time structure of the scintillation light. 
The experiment is presently under construction and commissioning on surface at CERN. Cryogenic operation and light detection performance was recently confirmed 
in a test run of the full 1\,ton liquid argon target under purely calorimetric 
operation and with a prototype light readout system. This note describes 
the experimental concept, the main detector components and presents some first results.
\end{abstract}

\vspace{-3mm}
\section{Introduction}
\vspace{1mm}

The best limit on spin-independent WIMP cross-sections is presently given by
the CDMS experiment\,\cite{cdms}, which records phonon and ionisation signals 
of particle interactions in cryogenic semiconductor targets\,\footnote{Total 
target mass is around 6\,kg} with high sophistication. 
A row of noble liquid dark matter experiments 
are currently striving for a similar kind of experimental perfection, actually 
starting to challenge these best limits\,\cite{xenon10}. 
While noble liquids are equally blessed with high scintillation 
and ionisation yields, they feature a high potential to be scaled up 
in the ton or even multi-ton target range, 
\begin{wrapfigure}{r}{0.4\textwidth}
\vspace{-6mm}
\centerline{\includegraphics[width=0.4\textwidth]{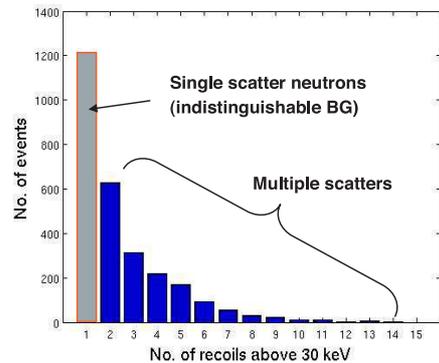}}
\vspace{-3mm}
\caption{Interaction multiplicity of background neutrons in \ardm\ (MC).}
\label{fig:nspect}
\vspace{-5mm}
\end{wrapfigure}
which makes this technology so popular and rapidly progres\-sing.
Large target sizes require very high radiopurity of the target, not only to
limit feedthrough of background in the nuclear recoil band, but also to 
comply with a maximal tolerable trigger rate. This is a main concern 
in the case for liquid argon (LAr) due to the long lived $^{39}$Ar isotope. 
Recent progress in liquifying underground gases and extracting substantial 
quantities of $^{39}$Ar depleted liquid argon gives however a 
promising outlook for this technology.
Self-shielding\,\footnote{against external radiation, above all gammas, but not neutrons},
one of the strongest motivations to go to 
large masses, is very much improving for larger and larger target sizes.  
Further on a large detector is  able to determine, on a statistical base,
the background of single scattering neutrons in the data from the 
multiplicity distribution of neutron interactions.
Figure\,\ref{fig:nspect} shows a frequency distribution of 
neutron interactions on the example of the \ardm\ 
geometry\,\cite{lili}.  
A hydrogen rich shield will nevertheless be used 
to protect the detector in its final working location from 
background neutrons (after a measurement of the spectrum).

In liquid argon, both, the scintillation light to charge ratio
and the temporal structure of the light emission\,\cite{kubot,suzuk,hitach}
can be used for electron- to nuclear-\,recoil discrimination. 
This is due to the ionisation density dependent ratio of the two argon excimer
ground states ($^{1}\Sigma^{+}_{u}$\,and\,$^{3}\Sigma_{u}^{+}$), 
which are responsible for the VUV luminescense of LAr. 
The large difference of their radiative lifetimes 
($\approx$\,10$^{3}$) allows for excellent recoil separation down to 
20\,\keve\ (electron equivalent scale)\,\cite{macK}.
This effect alone is used to classify recoils by some 
single phase experiments (e.g.\,DEAP/CLEAN,\,XMASS\,\cite{deapclean,xmass}). The two 
phase configuration (as we employ it in \ardm ) allows for an additional 
measurement of the ionisation charge (e.g.\,XENON,\,WARP,\,LUX\,\cite{xenon,warp,lux}\,experiments).
A drawback of the argon technology is the short wavelength of the scintillation light 
(128\,nm) and the already mentioned presence of the $^{39}$Ar $\beta$-emitter.
However, because of form factors, argon is less sensitive 
to the threshold of the nuclear recoil 
energy, than is e.g.~xenon. For the same reason the 
recoil energy spectra of argon and xenon are quite different. 
These liquids are therefore complementary in providing a 
crosscheck once a WIMP signal has been found.

\vspace{-2mm}
\section{Conceptual design}
\label{sect:design}
\vspace{1mm}

\ardm\,\cite{ardm} was projected as a ton scale liquid argon 
spectrometer aiming at the detection of nuclear recoils above 30\,\kevr\
(i.e.\,$\approx$\,10\,\keve ). At the same time it serves as a pro\-to\-type 
unit for future large LAr detectors, for more sensitive DM experiments or
next ge\-ne\-ration neutrino observatories. 
Three dimensional imaging and event by event interaction 
type identification will be used to reach high background suppression.
About 400 VUV photons and a couple of ionisation 
charges\footnote{if the electrical field is above 1\,kV/cm}
are typically produced in LAr by a WIMP interaction at 30\,keV.
The background rejection will be achieved by the combination of
cuts on the fiducial volume, the event topology (excl.~multiple 
scatter), the light to charge ratio and the 
temporal structure of 
\begin{wrapfigure}{r}{0.43\textwidth}
\vspace{-5mm}
\centerline{\includegraphics[width=0.4\textwidth]{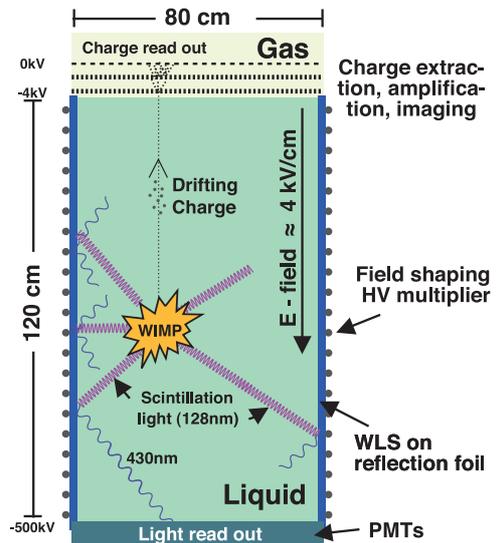}}
\vspace{-3mm}
\caption{Conceptual design of \ardm .}
\label{fig:concept}
\vspace{-5mm}
\end{wrapfigure}
the light emission.
The tech\-nical requirements comprise a large volume electric field, a large area 
position sensitive charge readout (3rd dimension from drift time), 
a fast and large area light readout and an efficient 
liquid argon purification system. The event trigger is generated from 
the light signal. Figure\,\ref{fig:concept} shows 
a sketch of the two-phase operating mode of the detector. 
In a particle interaction excited and ionised argon atoms form 
the argon excimer states\,\cite{suzuk} which decay under the 
emission of 128\,nm\,VUV radiation. This light 
can not be absorbed by neutral argon atoms and hence propa\-gates to the side 
walls of the detector which are coated with the wave shifting material 
tetra\-phenyl\-buta\-diene (TPB). VUV light is absorbed 
and with high efficiency re-emitted at around 430\,nm, the region 
of high quantum efficiency of 
borosilicate windowed bialkali PMTs.
By diffusive reflection on the side walls, the light is transported
to the bottom of the apparatus to an array of 14 hemispherical 8'' 
PMTs. The strong electric field is capable of preventing some
free electrons in the densely ionised region around a nuclear recoil 
from recombining and sweeps them to the surface of the liquid, where 
they are extracted into the gaseous phase of the detector.
The charges are then multiplied by means of a rigid large gas electron 
multiplier (LEM) and finally recorded by 
a position sensitive read out plane\,\cite{lem}.

\vspace{-4mm}
\section{The experimental components}
\vspace{1mm}

\begin{figure}[h]
\centerline{
\includegraphics[width=0.38\textwidth]{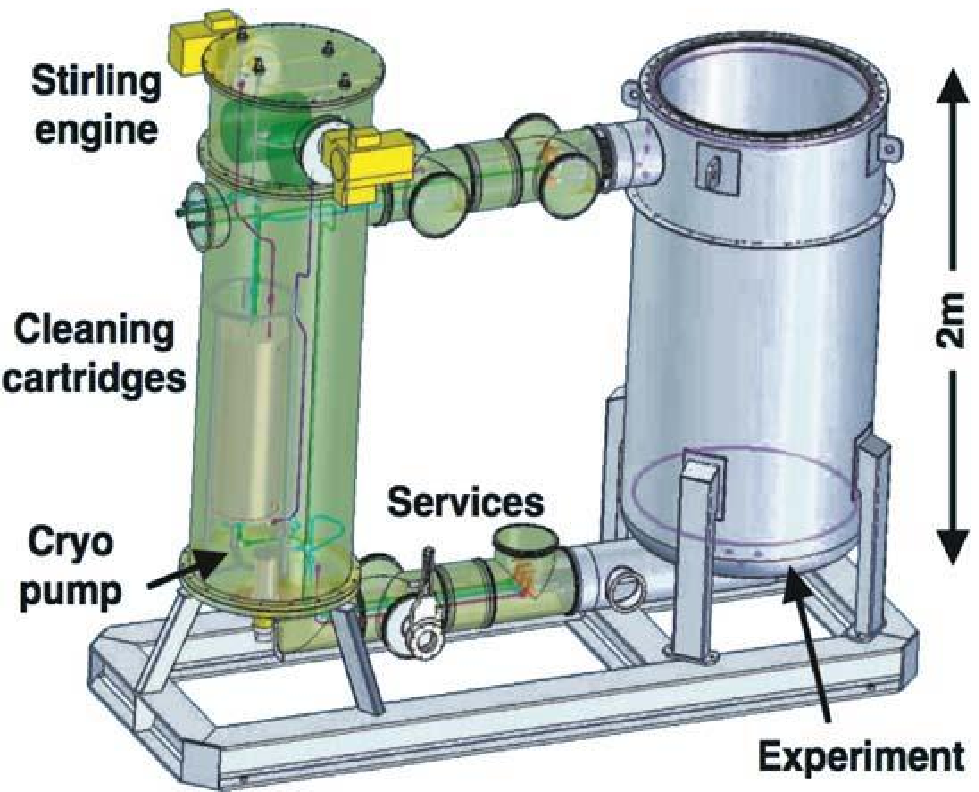}
\hspace{0.02\textwidth}
\includegraphics[width=0.24\textwidth]{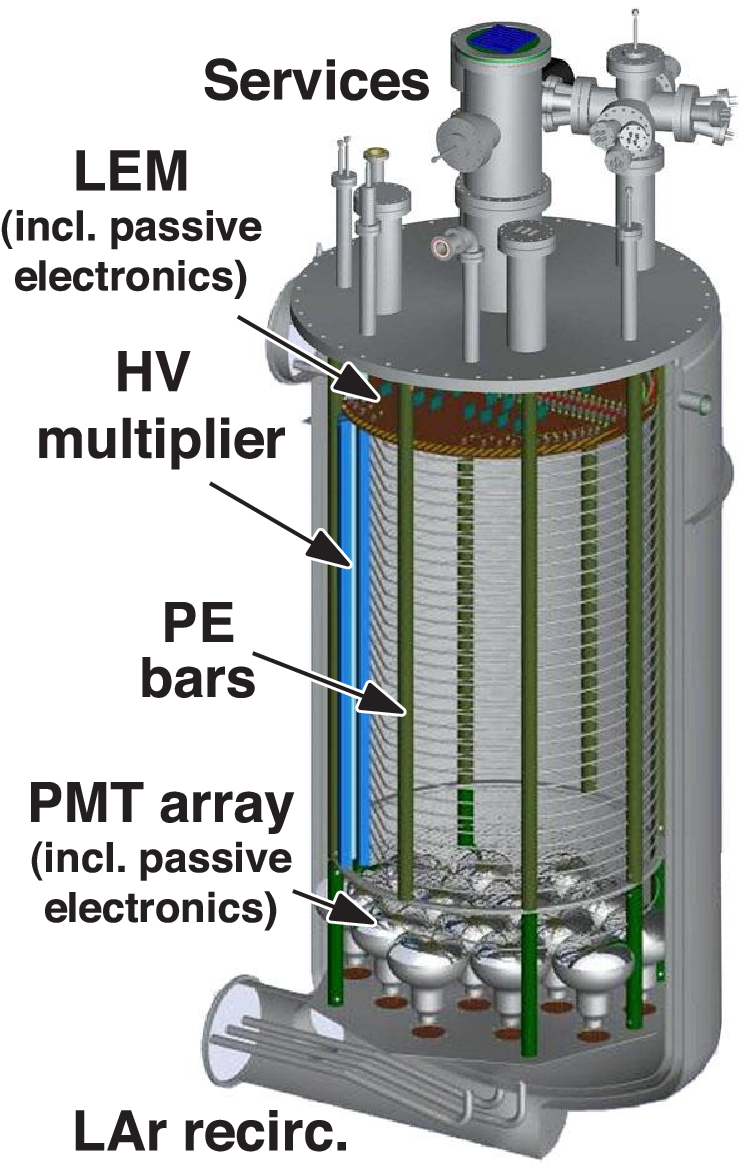}
\hspace{0.02\textwidth}
\includegraphics[width=0.30\textwidth]{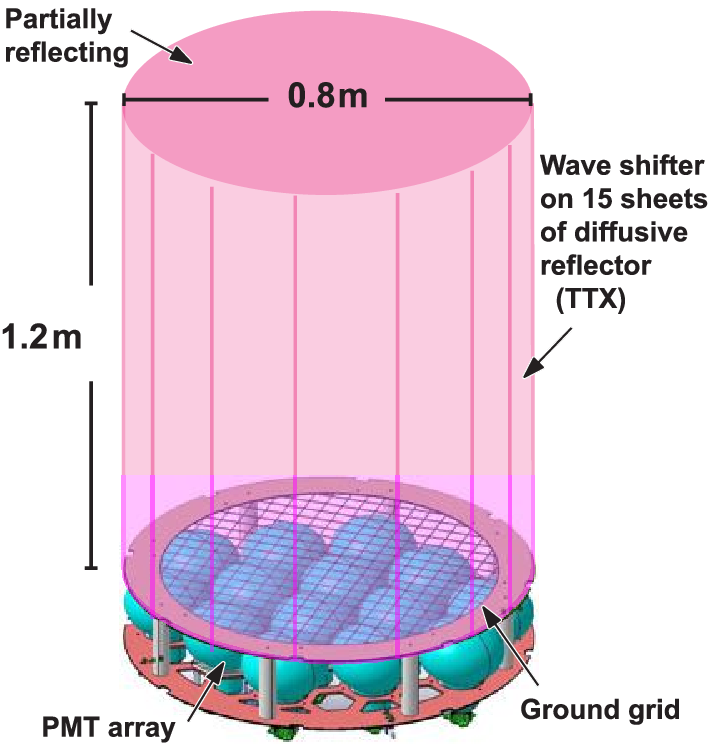}}
\caption{Left: cryo-system and main dewar; middle: detector 
components mounted to the top flange by polyethylene bars;
right: arrangement of the wavelength-shifter/reflector foils 
in light diffusion cell arrangement inside the field shaper rings.}
\label{fig:expdesign}
\vspace{-1mm}
\end{figure}

Figure\,\ref{fig:expdesign} left shows the mechanical arrangement of the 
cryogenic cooling and cleaning system of the setup 
together with the main stainless steel 
dewar (containing roughly 1800\,kg of liquid argon). 
An inner cylindrical volume of 80\,cm diameter and 120\,cm height 
is delimited by round ring electrodes (field shapers) constituting the
850\,kg active LAr target in a vertical TPC 
configuration (Fig.\,\ref{fig:expdesign} middle). 
The field shaper rings are connected to a 210 stage HV diode-capacitor 
charge pump system (Cockroft-Walton circuit) which is fully 
immersed in the
\vspace{-3mm}
liquid argon and designed to reach up to -500\,kV ($\approx$\,4\,kV/cm) at the bottom cathode.
Another unique feature of the experiment is the use of a LEM, which is placed 
5\,mm above the liquid level in the gas.
The positional readout is achieved by segmenting the upper LEM 
surface and the anode plate with 1.5\,mm wide $x$ and $y$-strips 
respectively. In total there are 1024 readout channels 
which are AC coupled to charge sensitive preamplifiers 
located externally on the top flange of the apparatus. 
Because the LEM is operated in very pure argon gas, which cannot 
quench charge avalanches, it has to be built with 
considerable attention to HV discharges.
To read scintillation light homogeneously over the full volume we 
adopted a (wave shifting) 
diffusion cell design with the PMT array at the bottom in the liquid.
This keeps the system simple\,\footnote{large 
area VUV sensitive photosensors, e.g.\,MgF$_2$ windowed PMTs
are commercially not available} and scalable.
The shifting of the VUV into the range of high quantum 
efficiency of the bialkali PMTs is done by
a thin layer ($\approx$1\mgcs) of tetra\-phenyl\-buta\-diene 
(TPB) evaporated onto 15, cylindrically arranged, 25\,cm wide reflector 
sheets which are located in the vertical electric field. 
These sheets, made out of the PTFE fabric Tetratex{\small\texttrademark} 
(TTX), are clamped to the upper- and lowermost field 
shaper rings. The PMTs (Hamamatsu R5912-02MOD-LRI) 
were made from particularly radiopure borosilicate glass
and are sensitive to single photons. They are equipped with with 
a Pt-underlay under their photocathodes for operation at LAr 
temperatures. This ineluctably 
reduces their quantum efficiency by roughly one third to a value between
15 and 20\%\,\cite{granada}.
The PMT glass windows are also coated with TPB to convert directly 
impinging VUV photons. The average number of recorded photoelectrons 
(n$_{\rm pe}$) is expected (laboratory measurements)
to be in the order of 1\,pe/keV$_{\rm ee}$.
The development of the light detection system 
and particularly the operation of gaseous argon test cells 
with \alp\ particle excitation were described 
in earlier work\,\cite{vuvidm,lumquench,wls}. 

\vspace{-2mm}
\section{First test run of \ardm\ on the full 1 ton LAr target}
\vspace{1mm}

A first run of the experiment was undertaken in May 2009 to test
cryogenic functionality (incl.~various safety installations) with a prototype
light read out system (7 PMTs out of 14 were mounted). No internal
electric field was yet present but background events and more over 
the response of the detector to external gamma sources could be
explored for the first time. 
\begin{wrapfigure}{l}{0.38\textwidth}
\vspace{-3mm}
\centerline{\includegraphics[width=0.34\textwidth]{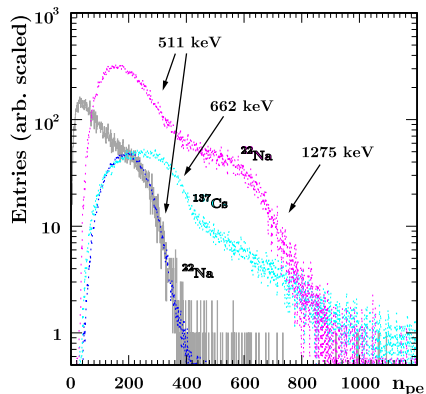}}
\vspace{-4mm}
\caption{gamma spectra in \ardm .}
\label{fig:gamspect}
\vspace{-4mm}
\end{wrapfigure}
Cryogenics proofed to work well, showing stable operation at high argon purity over 
the full 2 weeks of operation despite no liquid re\-cir\-cula\-tion and cleaning. 
LAr purity was monitored by the lifetime of the second scintil\-lation component 
(see\,\cite{lumquench} for details on this technique), recording a stable high 
value of $\approx$\,1500\,ns over the entire test period.
Figure\,\ref{fig:gamspect} shows the response of the detector to
external gamma sources, $^{137}$Cs (190\,kBq)  and $^{22}$Na (20\,kBq),
where the emission of the latter was controlled and triggered by an 
external crystal scintillator. The shoulders of full gamma ab\-sorption are 
marked with arrows. Dashed spectra stem from data triggered 
on large PMT signals at a threshold of roughly 150\,\keve\ 
($^{22}$Na data was taken in coincidence with the crystal).
The full, grey spectrum was recorded triggering on the external 
crystal only and shows an unbiased 511\,keV response (event reconstruction
down to 50\,keV$_{\rm ee}$ possible). We find an average
yield for this prototype light read out of $\approx$\,0.5\,pe/\keve , 
well in agree\-ment with the expected number of around 1\,pe/\keve\
of a fully equipped detector. A first comparison of the spectra with 
MC confirms an average value of 18\% for the light yield of the 
Pt-underlay PMTs at cold.

While R\&D work for sub detector parts is now finalising, main mechanical 
components are set together at CERN. Following a successful commissioning at 
surface we consider a deep underground operation. 

\vspace{-4mm}
\ack

This work is supported by grants from the Swiss National 
Science Foundation and ETH Z¬\"urich.

\vspace{-4mm}
 
\section*{References}
\vspace{2mm}



\begin{thebibliography}{99}

\bibitem{cdms}
Ahmed Z {\it et~al.} (CDMS collaboration) 2009 {\it Phys.\,Rev.\,Lett.} {\bf 102} 011301

\bibitem{xenon10}Angle J {\it et~al.} (XENON10 collaboration) 2008 
{\it Phys.\,Rev.\,Lett.} {\bf 100} 021303

\bibitem{lili}Kaufmann L 2008 {\it Doctoral Diss. ETH Z\"urich} No.\,17806

\bibitem{kubot}Kubota S {\it et~al.} 1978 {\it J.\,Phys.\,C} {\bf 11} 2645

\bibitem{suzuk}Suzuki M {\it et~al.} 1982 {\it NIM} {\bf 192} 565

\bibitem{hitach}Hitachi A {\it et~al.} 1983 {\it Phys.\,Rev.\,B} {\bf 27} 9 5279

\bibitem{macK}Lippincott W {\it et~al.} 2008 {\it Phys.\,Rev.\,C} {\bf 78} 035801

\bibitem{deapclean} DEAP/CLEAN: http://deapclean.org

\bibitem{xmass}Abe K {\it et~al.} 2008 {\it J.\,Phys.\,Conf.\,Ser.} {\bf 120} 042022

\bibitem{xenon} XENON100: http://www.physik.uzh.ch/groups/groupbaudis/xenon/

\bibitem{warp} WARP: http://warp.lngs.infn.it/papers/proposal.pdf

\bibitem{lux} LUX: http://www.luxdarkmatter.org

\bibitem{ardm}\ardm , Rubbia A 2006 {\it J.\,Phys.\,Conf.\,Ser.} {\bf 39} 129

\bibitem{lem} Badertscher A {\it et~al.} 2009 {\it NIMA A} doi:10.1016/j.nima.2009.10.011 


\bibitem{granada}Bueno A {\it et~al.} 2008  {\it JINST}  {\bf 3} P01006

\bibitem{vuvidm}Regenfus C 2007 {\it Proc.\,of IDM2006, Rhodes Greece, World Scientific}, 
page 325 

\bibitem{lumquench}Amsler C {\it et~al.} 2008 {\it JINST} {\bf 3} P02001

\bibitem{wls}Boccone V {\it et~al.} 2009 {\it JINST} {\bf 4} P06001




\end{thebibliography}
\end{document}